\documentstyle[11pt,paspconf,psfig]{article}

\begin{document}

\title{Elliptical Galaxies: Detailed Structure, Scaling Relations and
Formation}

\author{Ralf Bender \& Roberto P. Saglia}
\affil{Universit\"ats-Sternwarte, Scheinerstr. 1, D--81679 Munich,
Germany}



\begin{abstract}
The last decade of research on elliptical galaxies has produced a
wealth of new information concerning both their detailed structure and
their global scaling relations. We review the old and new results about
isophote shapes and subcomponents
(Sect. \ref{isophotes}), scaling relations of global
parameters and redshift evolution 
(Sect. \ref{scaling}), and the ages, metallicities
(Sect. \ref{populations}) and abundance ratios
(Sect. \ref{abundances}). Finally, we confront the observations with
hierarchical formation scenarios of elliptical galaxies
(Sect. \ref{formation}).

The picture
emerging from this variety of observational evidence is broadly
consistent with the merging scenario of hierarchical structure
formation models, but the stellar population properties of ellipticals
pose some challanges. The formation of ellipticals must have always
involved some dissipation. The gas fraction at the last major merger
event presumably has strong influence on their present day properties. 
 Most elliptical galaxies are old systems, but disky
ellipticals might be younger than boxy objects and have more extended
star formation histories. The small scatter and the redshift
variations of the scaling relation are compatible with passive
evolution. The high Mg/Fe overabundances of luminous (boxy) ellipticas
point to a rapid star formation episode, while low-luminosity objects
have values of Mg/Fe nearly solar, allowing  for an extended
star formation history.
\end{abstract}


\section{Isophote shapes, sub-components and kinematics}
\label{isophotes}

Isophote shapes have proven to be an easily measurable indicator for
the internal kinematics and core structure of ellipticals as well as
for their radio and X-ray properties (e.g., Bender 1988a, 1997, Bender
et al. 1989, Faber et al. 1997, Beuing et al. 1998).  Boxy ellipticals are
characterized by anisotropy and shallow cores, and by stronger than
average radio and X-ray emission. Disky ellipticals contain faint
stellar disks, are rotationally flattened, have power-law inner density
profiles and show little or no radio and X-ray emission. 

The disk-to-bulge ratios of disky ellipticals can be as high as 0.3
and overlap those of S0-galaxies (Rix \& White 1990, Scorza \& Bender
1995).  Generally, the spheroids are rotationally supported and the
angular momenta of disks and bulges are parallel to each other
indicating that the disks were not randomly accreted at late times
(Bender et al. 1993, Scorza \& Bender 1995). The high density
power-law centers of disky Es (Faber et al. 1997) and the fact these
galaxies do indeed harbour disks show that dissipation was essential for their
formation. Disky
Es simply seem to form the extension of the Hubble sequence to the
lowest disk-to-bulge ratios (Bender 1988a, 1990, Kormendy \& Bender
1996).

\begin{figure}
\psfig{figure=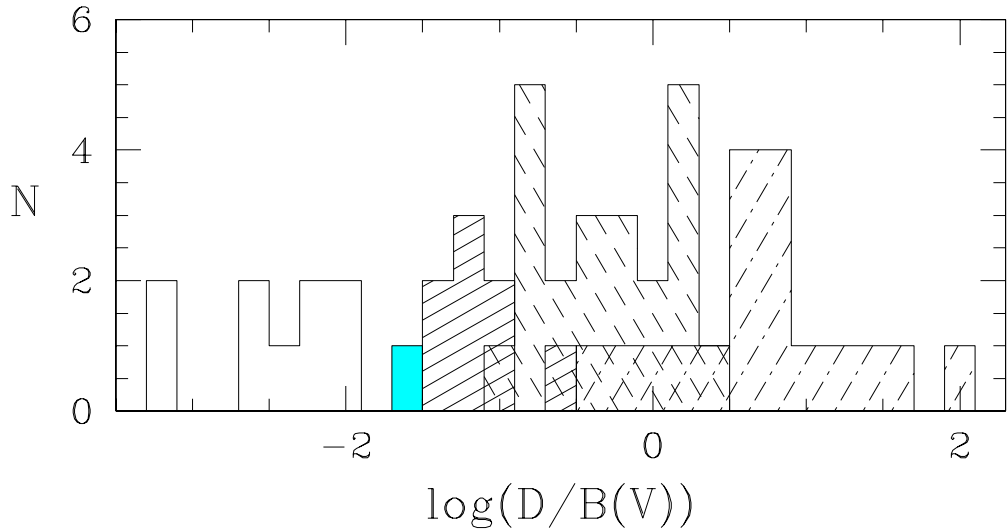,width=\linewidth}
\caption{The histogram of the D/B ratios of the (kinematically
decoupled) disks of {\it boxy} galaxies, of {\it disky}
galaxies (full-line shaded), of S0 galaxies (dash-line shaded), and of
spiral galaxies (dot-dash line shaded). The position of NGC 5322 is
indicated by the grey colour (from Mehlert 1998). } 
\label{figdbhist}
\end{figure}

The kinematic structure of boxy Es is generally more complex than the
one of disky Es.  Peculiar velocity fields, like intrinsic minor axis
rotation and kinematically decoupled central regions (Franx \&
Illingworth 1988, Jedrezjewski \& Schechter 1988, Bender 1988b) are
frequently found in boxy Es, almost never in disky Es. Intrinsecally peculiar
velocity fields are a natural by-product of merging of star-dominated
systems (e.g. Hernquist \& Barnes 1991; note that projected velocity fields
of triaxial bodies can also show peculiarities, see Statler
1994). Unlike shells or ripples (e.g., Schweizer 1990), these features
are long-lived and carry 'genetic' information about the formation
process of the main (i.e. inner) parts of the galaxy. Kinematic
decoupling is generally caused by flattened, rapidly rotating, disk-
or torus-like components that dominate the light in the central few
hundred pc to kpc (Bender 1990, Rix \& White 1992, 
Surma \& Bender 1995, Mehlert et al. 1998). Their
masses are of the order 10$^9$ to 10$^{10}$ M$_\odot$, i.e. only
massive mergers can provide enough gas for their formation. The
decoupled components do however contribute always less than a few
percent to the total light. In fact, the ratio of
decoupled-light-to-total-light overlaps with the disk-to-total-light
ratios of disky ellipticals at values of about 0.03 
(see Fig. \ref{figdbhist}).

Motivated by this finding one can indeed arrange  all galaxies in
a disk-to-total sequence ranging from spirals and S0s via disky
ellipticals to boxy ellipticals. Towards lower disk-to-total ratios
the structures of the galaxies become more and more pressure supported
and the anistropy of the spheroidal components may increase as well. 
Finally, around very low disk-to-total ratios,
kinematical decoupling between disks and spheroids sets in. In the
most extreme cases, no disk component is formed at all.

\section{Scaling relations and global homogeneity}
\label{scaling}

Ellipticals define a two-dimensional manifold in the three-dimensional
space of their global structural parameters (effective radius $R_e$,
mean effective surface brightness $<SB>_e$, velocity dispersion
$\sigma$), the so-called Fundamental Plane (FP, Djorgovski
\& Davis 1987, Dressler et al. 1987). Its defining relation is 
$\log R_e = 1.25 \log\sigma + 0.32 <{\rm SB}>_e + const.$ (e.g. J\o
rgensen et al. 1996). It seems to be independent from environment
(Burstein et al. 1988, J\o rgensen et al. 1996) 
and is valid for S0s and, with slight
changes, for dwarf ellipticals, too (Bender, Burstein \& Faber
1992). It is now generally agreed that the FP is simply
a consequence of the Virial theorem and the fact that E galaxies have
similar mass-to-light ratios and close to homologous structure at a
given luminosity (e.g. Faber et al. 1987, Djorgovski et al. 1989,
Bender, Burstein \& Faber 1992, but see Pahre et al. 1998a, 1998b). 

It is very remarkable that, despite of the large variety in internal
dynamics and structure, the scatter perpendicular to the FP
is as small as observed. Jorgensen et al. (1996) find a typical
rms-scatter of 20\% in $R_e$ or $M/L$. In the case of the Coma
cluster core, the scatter is smaller than 10\% (Saglia et al. 1993, Mehlert
1998).  This yields tight constraints on variations in the
initial stellar mass function, in dark matter contributions,
metallicities and ages, see Renzini \& Ciotti (1993), Ciotti et
al. (1996), Graham \& Colless (1997).  Another indicator for the
relative homogeneity of ellipticals on the global level is that the
objects have close to isotropic velocity distributions (Rix et
al. 1997, Gerhard et al. 1998, Kronawitter et al., this conference) and are
only mildly triaxial with mostly near oblate shapes (e.g. Franx et
al. 1991). 
This is significantly different from the preferred
prolate-triaxial shape of dark matter halos formed in cosmological
N-body simulations (e.g. Frenk et al. 1988).

One explanation for the surprising homogeneity of ellipticals on the
global level could be the presence of centrally concentrated gas in
the merging events Es underwent during their formation. Of similar
importance may have been supermassive black holes or steep central
density cusps (Gerhard \& Binney 1985, Dubinski 1994, Valluri \& 
Merritt 1998). Any
sufficiently concentrated central mass will depopulate box orbits in
favor of z-axis tubes (Barnes \& Hernquist 1996).  Box orbits are the
backbone of triaxial objects (see e.g. Merritt 1997, Merritt \&
Quinlan 1998), while z-tubes
are dominant in rotationally flattened objects. Scattering of box
orbits would preserve disks and decoupled central components but would
isotropize the velocity distribution and reduce the triaxiality of the
galaxy. If gas fractions or black hole masses are similar for
progenitors of similar mass then Es of similar luminosity may have
largerly similar phase space structure despite of decoupled cores or
disk components. 
These features are indeed mostly
caused by axisymetric components which will not be destroyed by orbit
scattering.

\begin{figure}
\begin{centering}
\psfig{figure=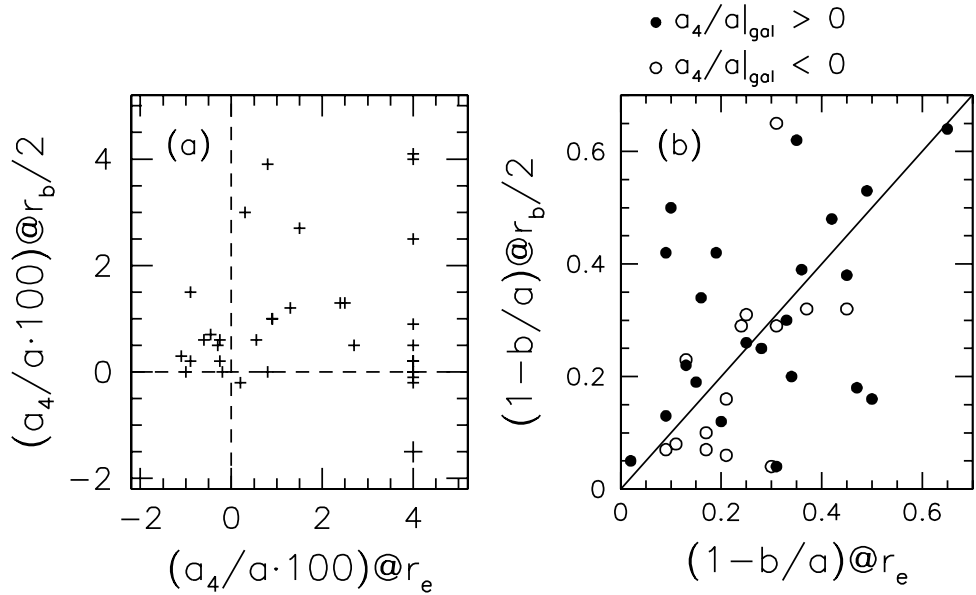,width=\linewidth}
\end{centering}
\caption{(a) The comparison between the $a_4$ parameter averaged
over $r_e$ and $r_b/2$. (b) The comparison between the ellipticity
$1-b/a$ averaged over $r_e$ and $r_b/2$ (from Tymann 1998). Isophotes
inside half the break radius are elliptical or {\it disky} and no more
flattened than the outer regions. } 
\label{figa4e}
\end{figure}

The importance of orbit scattering can be tested by analysing the
photometric structure of the innermost regions of elliptical galaxies
and comparing it with the one of the main bodies. Orbit scattering should make
the inner isophotes rounder and, if boxiness is caused at least in
part by box orbits, also more elliptical or disky (Valluri \& Merritt
1998). In an analysis of
WFPC2 images of elliptical and S0 galaxies from the HST archive, 
preliminary results by
Tymann (1998) show that isophotal shapes of Es and S0s indeed do
change around the photometric break radius (for a discussion of 
break radius and inner light profiles see Kormendy's
contribution to this conference). At half the break radius, none of the
resolved galaxies is boxy anymore, the isophote shapes are either
elliptical or disky (Figure \ref{figa4e}). However, isophotes do not become
less flattened towards smaller radii. These findings
are in part compatible with orbit scattering, but more
data and simulations are needed to draw even preliminary conclusions.

\section{Stellar Populations and Ages}
\label{populations}

At any given mass, the stellar populations of elliptical galaxies are
remarkably homogenous, consistent with the small scatter about the
FP. Colors and line-strengths are generally one-to-one
correlated and scale with luminosity or, even more tightly, with
velocity dispersion $\sigma$ (e.g.  Burstein et al. 1988, Dressler et
al. 1987, Bower et al. 1992, Bender, Burstein \& Faber 1993). It is
important to note that there is no difference in the Mg$-\sigma$
relation between disky and boxy Es or between kinematically peculiar
and regular Es.  However, there are hints for a weak dependence of the
color$-\sigma$ and Mg$-\sigma$ relations on the presence of rather
short-lived peculiarities (due to accretion of younger stars,
Schweizer et al. 1990) and, possibly, on environmental density
(Guzm\'an \& Lucey 1992, J\o rgensen 1997), but generally 
there is also no significant evidence for a population
difference between field and cluster ellipticals of similar velocity
dispersion (Colless et al. 1998, Bernardi et al. 1998).

The fact that the Mg-index and colors increase and population
gradients steepen with luminosity (e.g. Carollo et al. 1993, Mehlert
et al. 1999a, 1999b) are further strong evidence that a merging
picture without dissipation does not work.

\begin{figure}
\begin{centering}
\psfig{figure=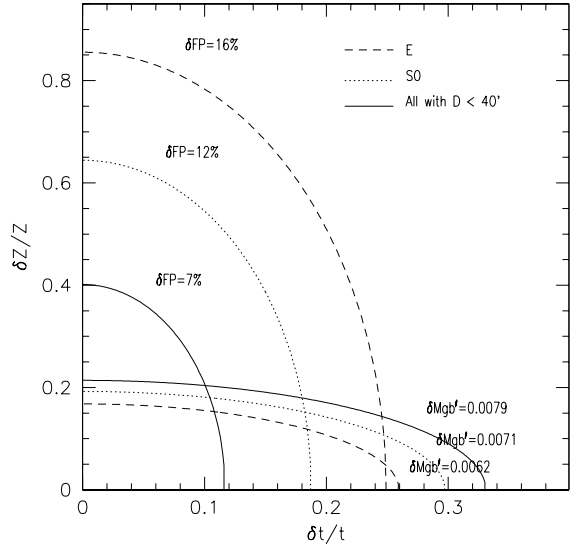,width=8cm}
\end{centering}
\caption{The global constraints on the scatter in age and metallicity
from the combined scatter of the Mg-$\sigma$ and FP relation 
(from Mehlert 1998).} 
\label{figfpmg}
\end{figure}

Using stellar population synthesis models (e.g., Worthey 1994)
one can estimate the combined scatter in age and metallicity from the
observed scatter in the color$-\sigma$ or Mg$-\sigma$ relations
(e.g. Bower et al. 1992; Bender, Burstein \& Faber 1993). 
The strongest constraints however follow from a combined analysis of the
scatter around the FP and the Mg-$\sigma$-relation. 
Colless et al. (1998) note that while the FP scatter is most
sensitive to the scatter in age, the Mg-$\sigma$ scatter is most
sensitive to the scatter in metallicity of the stellar
populations. Figure \ref{figfpmg} (Mehlert 1998, Mehlert et al. 1999a,
1999b) shows how the combined small scatters of the
Mg-$\sigma$ and FP of the Coma cluster early-type galaxies allow for
no more than $\approx $20 \% scatter in relative metallicity and age
at a given $\sigma$. The constraint is even tighter for the sample of
galaxies of the core of the cluster, where $\delta t/t\approx 10$\%.

This is consistent with the current observations of elliptical galaxy
evolution as a function of redshift.  The color-magnitude relation
(Standford et al. 1998) up to $z\approx 1$, the Mg-$\sigma$-relation
(Bender et al. 1996, Ziegler \& Bender 1997) up to $z\approx 0.4$, and
the FP relation (van Dokkum \& Franx 1996, Kelson et al. 1997, Bender
et al. 1998, van Dokkum et al. 1998) up to $z\approx 0.8$ all seem to
be compatible with {\it passive} evolution, implying high formation
redshifts.

The age constraints for lower luminosity Es, which mostly belong to
the disky class, are less tight due to smaller samples or larger
scatter in Mg and colors at smaller $\sigma$.  In fact, it is
indicated that low-luminosity Es ($\rm M_T \approx -18$) seem to be
systematically younger than giant Es ($\rm M_T \approx -21$), see
Faber et al. (1995) and Worthey (1996). Note that this trend runs
opposite to the one expected in a cold-dark-matter model (Kauffmann et
al. 1997, Kauffmann 1998, Baugh et al. 1998). 
The apparently smaller ages of low
luminosity Es could in fact be caused by the faint disks they
contain. These disks may become more dominant towards lower
luminosities and may have had extended star formation histories. Hints
for this have been found by de Jong \& Davies (1997).

\section{Abundance Ratios and Star Formation Time Scales}
\label{abundances}

Another way to extract information about the star formation history of
ellipticals is to analyse their element abundance ratios.  For {\it
luminous} Es, Worthey et al. (1992), Davies et al. (1993), Mehlert et
al. (1998) and others found consistently that Mg is overabundant
relative to Fe. Over a larger luminosity range, [Mg/Fe] seems to be
correlated with velocity dispersion: faint Es have [Mg/Fe]$ \approx
0$, while luminous Es reach [Mg/Fe]$\approx 0.4$ (Gonzalez 1993,
Fisher, Franx \& Illingworth 1995). Furthermore, Paquet (1994) could
show that, in luminous Es, other light elements like Na and CN are
overabundant relative to Fe as well. Within the galaxies, the [Mg/Fe]
overabundance is usually radially constant up to at least the
effective radii (Davies et al. 1993, Mehlert et al. 1998, Paquet
1994).  Generally, no distinction between 'normal' luminous Es and Es
with kinematically decoupled cores is indicated.

These results imply that the enrichment history of luminous Es differed
significantly from the one of the solar neighborhood, see e.g.,
Matteucci \& Greggio (1986), Truran \& Burkert (1995), Faber et
al. (1995), Worthey (1996).

Evidently, the enrichment of {\it massive} (high velocity dispersion)
Es was dominated by Supernovae II, as only they can produce a light
element overabundance.  Supernovae Ia basically just provide iron peak
elements (see, e.g., Truran \& Thielemann 1986). Because the yields of
SNII integrated over a plausible IMF result in [Mg/Fe]$\approx
0.3\,$dex at most (see Thomas, Greggio \& Bender 1998a), we can
conclude that the contribution of SNIa to the enrichment of the most
massive Es must have been small.

\begin{figure}
\begin{minipage}{0.49\linewidth}
\psfig{figure=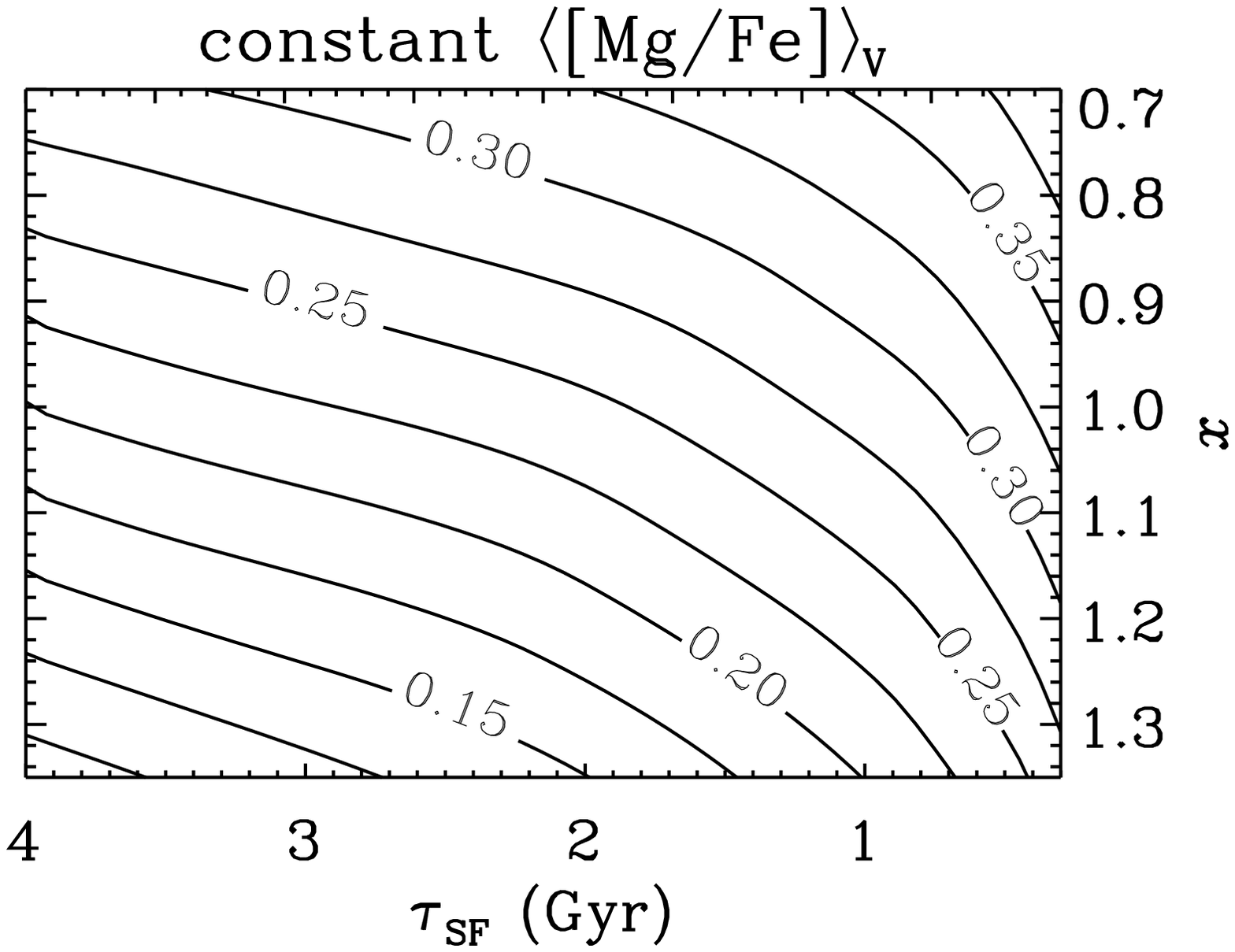,width=\linewidth}
\end{minipage}
\begin{minipage}{0.49\linewidth}
\psfig{figure=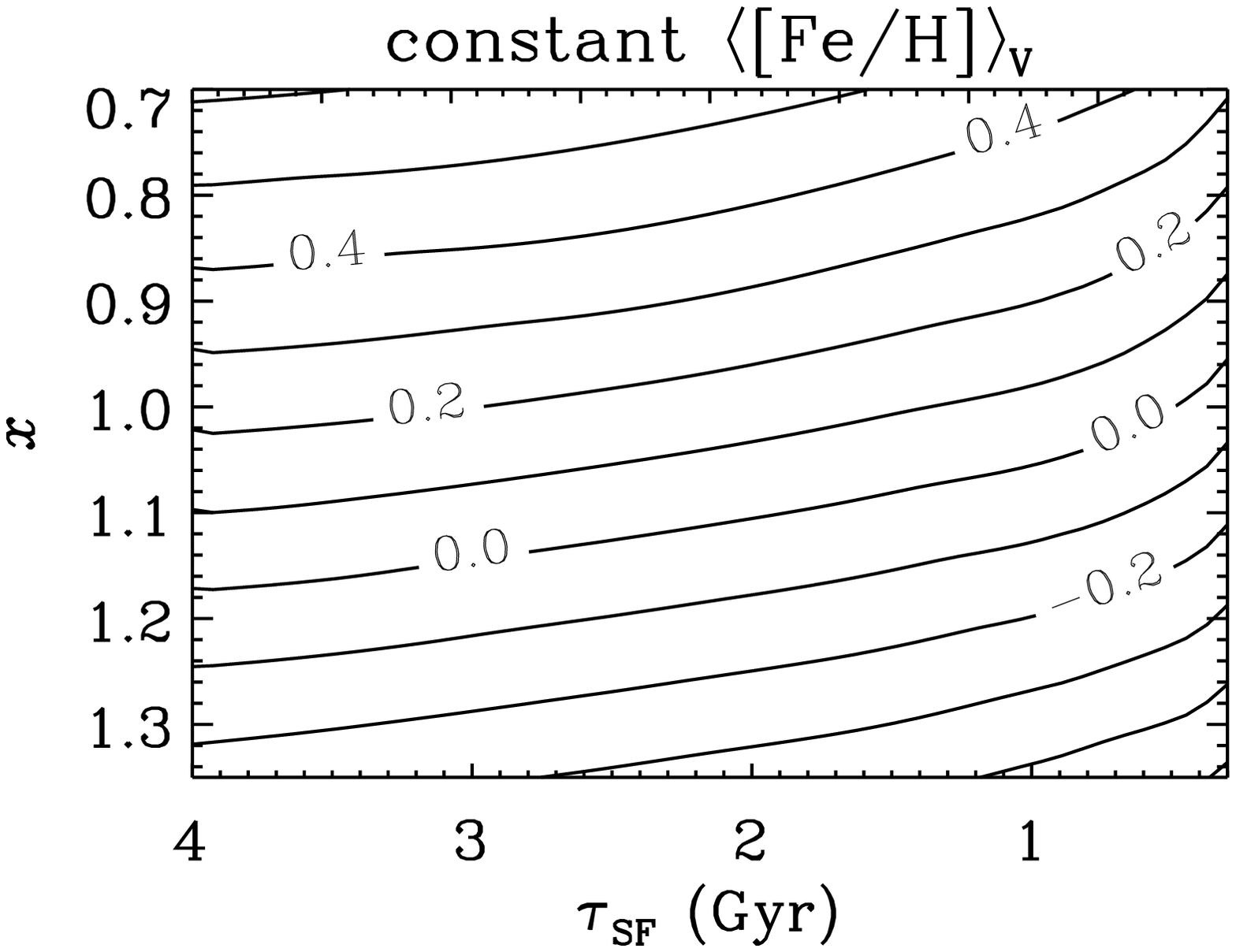,width=\linewidth}
\end{minipage}
\caption{The contours of constant $\langle[$Mg$/$Fe$]\rangle_V$ (left)
and of constant $\langle[$Fe$/$H$]\rangle_V$ (right) as a
function of the star formation time-scale $\tau_{SF}$ and the slope of
the IMF. Long star formation time scales cannot be reconciled with the
high Mg/Fe overabundance pbserved in elliptical galaxies even with a
modestly flat IMF (from Thomas et al. 1998b).}
\label{figthomas}
\end{figure}

The prevalence of Supernovae II and in turn the light element
overabundance in massive Es can have the following reasons:
(a) a star formation time scale smaller than about 1Gyr (SNI explode
in significant numbers only after a few times 10$^8$yrs after star
formation started, e.g. Truran \& Burkert 1995), (b) a top heavy
initial mass function, (c) a reduced frequency of binary stars
(leading to fewer SNI events). -- Option (c) is
rather unlikely because one expects the binary frequency to be
determined by the local process of star formation rather than by
global galaxy properties. In addition, a low binary fraction may be
inconsistent with the observed frequency of discrete X-ray sources in
old populations.  Neither does option (b) work well, because the
overabundance in massive Es reaches [Mg/Fe]$\approx 0.4\,$dex
(as is also observed in Galactic halo stars, Fuhrmann et
al. 1995). For such high overabundances, a flat IMF alone cannot solve
the overabundance problem. So, option (a), i.e. a short star 
formation time scale, seems to be necessary in any case.
In their recent study, Thomas, Greggio \& Bender (1998b) show
that the star formation time scale in massive Es was probably
shorter than about 1Gyr. Using a shallower IMF this time can be
somewhat extended but gets in conflict with the observed
abundance of iron in most ellipticals, which is about solar 
(see Fig. \ref{figthomas}).

Note that these considerations do not only apply to the cores of
luminous Es but for the bulk of their stars, since the [Mg/Fe]
overabundance is similar at all radii (see above).  And another
important conclusion can be drawn from these findings: since most
present day spirals have gas-to-star ratios smaller than 0.2 and disk
stars show solar element ratios, merging of objects similar to
present-day spirals cannot produce objects similar to most present-day
massive Es.

Since lower luminosity ellipticals have smaller light-element
overabundances, their star formation time scales are not severely
constrained. In fact, solar element abundance ratios could be
taken as a hint for extended star formation histories in smaller Es.

\section{Observations {\it vs.} hierarchical galaxy formation}
\label{formation}

In hierarchical galaxy formation scenarios galaxies are expected to
form via a sequence of merging and accretion processes (e.g. White
1993).  Merging is a key driver of star formation depleting the gas
with increasing galaxy mass (e.g. Katz 1992, Steinmetz \& M\"uller
1994, Kauffmann 1998, Baugh et al. 1998). Mergers with mostly stellar
progenitors are supposed to form ellipticals and are more likely to
occur if the progenitors are more massive. Note that the stellar
population properties (e.g. Mg-$\sigma$-relation, Mg-gradients etc.)
require that some gas must have been involved in the formation
of basically all ellipticals (see above).

The amount of gas present during a merger can have significant
influence on the dynamical structure of the merger product. As
discussed above, centrally concentrated gas, likewise density cusps or
black holes, will depopulate box orbits. For
some objects the gas fraction or central density concentration will be
so low that a `phase transition' from parallelized angular momenta of
disky and spheroidal components to kinematical decoupling can
occur. Simply speaking, less concentration implies a more exciting end
product.  In this scenario, disky Es are merger products as well but a
higher gas fraction helped to produce objects with properties similar
to what is expected from a {\it single} dissipative collapse
(e.g. Larson 1975).  Consistent
with this scenario, the vast majority of boxy Es (except for a small
sub-class of boxy companions to bright galaxies) has luminosities
above $L_*$, while the luminosity function of disky Es resembles the
one of S0s (Bender et al. 1993). As to be expected in hierarchical
galaxy formation, there is a variety of paths to form objects of
similar mass and this is why ellipticals of similar luminosity can
show quite different detailed properties. {\it This fact also implies that
for any given elliptical, isophotal shape may be a better indicator for
the formation history than luminosity.}

The low-redshift analogue of the late formation-phase of boxy Es may
be found in ultraluminous IRAS mergers (Schweizer 1990, Kormendy \&
Sanders 1992, Bender \& Surma 1992). Numerical simulations with large
particle numbers indicate that dissipationless violent relaxation
alone may indeed create boxy main bodies (Steinmetz 1995).
Furthermore, Bekki \& Shioya (1997) found indications
that the rapidity of the gas consumption during the merger may
influence the isophotal shape of the remnant. In their simulations,
rapid star formation produced boxy isophotes, slow star formation
disky isophotes. A large fraction of the molecular gas of the
progenitors concentrates in the central kiloparsec and can form a
kinematically decoupled center (Hernquist \& Barnes 1991, Barnes
1996). Indeed, the masses and metallicities of kinematically decoupled
centers in Es are quite similar to those of the central gas tori in
IRAS mergers (Sanders, Scoville, Soifer, 1991; Bender \& Surma 1992, Mehlert et
al. 1998).  We can speculate that very gas-rich progenitors (and/or
slow star formation during the merger (Bekki \& Shioja 1997) may also
create disky Es, especialy if gas falls in at late times from large
radii (Hibbard \& van Gorkom 1996). The analogy of E galaxy formation and
the IRAS merging process is unlikely to be perfect.  Especially, it
does not necessarily imply that many field Es formed in spiral-spiral
mergers at low redshift -- merging of any star-dominated progenitors
at any redshift may have produced similar remnants.

While the structural properties of ellipticals naturally arise in a
hierarchical model of galaxy formation, the stellar population
properties of ellipticals pose more of a challenge.  The generally
high overabundances of light elements over Fe in massive ellipticals
indicate very short star formation time scales and are in conflict
with forming a significant fraction of field ellipticals in late
mergers. Furthemore, the fact that low luminosity ellipticals appear
to be younger than high luminosity objects needs a convincing
explanation.

There is still the other possibility that current
stellar population models do not explain absorption indices accurately
enough to determine overabundances with sufficient reliability.
If the true overabundances in massive ellipticals are significantly
lower than current estimates, then this relaxes the observational
limit on more extended star formation at lower redshifts
significantly.

\acknowledgments

This work was supported by the Deutsche Forschungsgemeinschaft via SFB
375.  R.B. acknowledges travel support from the conference organizers
and the Max-Plank Gesellschaft.  We are grateful to J. Beuing,
D. Mehlert, L. Greggio, D. Thomas for discussions.

\end{document}